\PassOptionsToPackage{bookmarks=false}{hyperref}
\documentclass[sigconf]{acmart}

\usepackage{subcaption}
\usepackage{multirow}
\usepackage{listings}
\usepackage{color}
\definecolor{pblue}{rgb}{0.13,0.13,1}
\definecolor{pgreen}{rgb}{0,0.5,0}
\definecolor{pred}{rgb}{0.9,0,0}
\definecolor{pgrey}{rgb}{0.46,0.45,0.48}
\AtBeginDocument{%
  \providecommand\BibTeX{{%
    \normalfont B\kern-0.5em{\scshape i\kern-0.25em b}\kern-0.8em\TeX}}}

\copyrightyear{2020}
\acmYear{2020}
\setcopyright{acmcopyright}\acmConference[ASE '20]{35th IEEE/ACM International
Conference on Automated Software Engineering}{September 21--25, 2020}{Virtual
Event, Australia}
\acmBooktitle{35th IEEE/ACM International Conference on Automated Software
Engineering (ASE '20), September 21--25, 2020, Virtual Event, Australia}
\acmPrice{15.00}
\acmDOI{10.1145/3324884.3416578}
\acmISBN{978-1-4503-6768-4/20/09}

\begin{document}

\title{Retrieve and Refine: Exemplar-based Neural \\ Comment Generation}

\author{Bolin Wei}
\affiliation{%
  \institution{Key Lab of High Confidence Software \\ Technology, MoE (Peking University)\\ Beijing, China}
}
\email{bolin.wbl@gmail.com}

\author{Yongmin Li}
\affiliation{%
  \institution{Key Lab of High Confidence Software \\ Technology, MoE (Peking University)\\ Beijing, China}
}
\email{liyongmin@pku.edu.cn}

\author{Ge Li}
\authornote{Corresponding authors.}
\affiliation{%
  \institution{Key Lab of High Confidence Software \\ Technology, MoE (Peking University)\\ Beijing, China}
}
\email{lige@pku.edu.cn}

\author{Xin Xia}
\affiliation{%
  \institution{Faculty of Information Technology \\Monash University, Australia}
}
\email{xin.xia@monash.edu}

\author{Zhi Jin}
\authornotemark[1]
\affiliation{%
  \institution{Key Lab of High Confidence Software \\ Technology, MoE (Peking University)\\ Beijing, China}
}
\email{zhijin@pku.edu.cn}
\renewcommand{\shortauthors}{Wei, et al.}

\begin{abstract}
Code comment generation which aims to automatically generate natural language descriptions for source code, is a crucial task in the field of automatic software development. Traditional comment generation methods use manually-crafted templates or information retrieval (IR) techniques to generate summaries for source code. In recent years, neural network-based methods which leveraged  acclaimed encoder-decoder deep learning framework to learn comment generation patterns from a large-scale parallel code corpus, have achieved impressive results. However, these emerging methods only take code-related information as input. Software reuse is common in the process of software development, meaning that comments of similar code snippets are helpful for comment generation. Inspired by the IR-based and template-based approaches, in this paper, we propose a neural comment generation approach where we use the existing comments of similar code snippets as exemplars to guide  comment generation. Specifically, given a piece of code, we first use an IR technique to retrieve a similar code snippet and treat its comment as an exemplar. Then we design a novel seq2seq neural network that takes the given code, its AST, its similar code, and its exemplar as input, and leverages the information from the exemplar to assist in the target comment generation based on the semantic similarity between the source code and the similar code. We evaluate our approach on a large-scale Java corpus, which contains about 2M samples, and experimental results demonstrate that our model outperforms the state-of-the-art methods by a substantial margin.
\end{abstract}



\maketitle

\section{Introduction}
\label{sec:intro}
Code comments provide a clear natural language description for a piece of the source code, which can help software developers understand programs quickly and correctly~\cite{DBLP:conf/kbse/SridharaHMPV10}. Previous studies showed that during software maintenance, program comprehension takes more than half of the time~\cite{DBLP:journals/tse/XiaBLXHL18, DBLP:journals/tse/KoMCA06, Fjeldstad1982ApplicationPM, DBLP:journals/ibmsj/Corbi89}. Although proper comments are very helpful for software maintenance, they are absent or out-dated in many software projects~\cite{DBLP:conf/sigdoc/SouzaAO05}. On the other hand, manually writing comments is very time-consuming and labor-intensive, and the comments should be updated as the software is upgraded. Therefore, automatic comment generation becomes greatly crucial for software development and maintenance.

Creating manually-crafted templates is a common way to generate comments automatically~\cite{DBLP:conf/kbse/SridharaHMPV10,DBLP:conf/iwpc/MorenoASMPV13}. These methods defined different templates for different types of programs to generate readable text descriptions. Sridhara et al.~\cite{DBLP:conf/kbse/SridharaHMPV10} used Software Word Usage Model and heuristics to select important code statements, defined templates for each code statement, and generated corresponding comments. Moreno et al.~\cite{DBLP:conf/iwpc/MorenoASMPV13} predefined heuristic rules to extract information from source code, and defined templates for different types of information to help generate code summaries. Manually-crafted templates are introduced in these approaches to extract key information in the source code into comments, helping improve the readability and comprehensibility of comments. However, defining a template is a time-consuming task and requires extensive domain knowledge. Also, different projects might use different kinds of templates. 

As an alternative, information retrieval (IR) techniques are widely used in automatic comment generation~\cite{DBLP:conf/wcre/HaiducAMM10, DBLP:conf/iwpc/EddyRKC13, DBLP:conf/wcre/WongLT15, DBLP:conf/kbse/WongYT13}. Some researchers used IR techniques to select terms from source code to generate term-based comments~\cite{DBLP:conf/iwpc/EddyRKC13,DBLP:conf/wcre/HaiducAMM10}. Haiduc et al.~\cite{DBLP:conf/wcre/HaiducAMM10} applied the Vector Space Model and Latent Semantic Indexing to retrieve the appropriate terms, while Eddy et al.~\cite{DBLP:conf/iwpc/EddyRKC13} introduced a hierarchical topic model for comment generation. Based on the idea that software reuse is common~\cite{DBLP:journals/tse/KamiyaKI02, DBLP:conf/sigsoft/KimSN05},  other researchers leveraged code clone detection techniques to detect similar code snippets and used their corresponding comments for comment generation. Note that similar code snippets can be retrieved from existing open-source software repositories in GitHub or software Q\&A sites such as Stack Overflow~\cite{DBLP:conf/wcre/WongLT15, DBLP:conf/kbse/WongYT13}. However, code snippets may contain some information that is inconsistent with the content in comments of their similar code snippets. 

In recent years, more and more researchers have focused on applying neural machine translation models for comment generation and viewed the process of generating comments from source code as a language translation task (e.g., translating English to German)~\cite{DBLP:conf/iwpc/HuLXLJ18, DBLP:conf/acl/IyerKCZ16, DBLP:conf/icse/LeClairJM19}. These research works have adopted the mainstream encoder-decoder framework of neural machine translation, with source code as input and comments as output, and achieved state-of-the-art performance on the comment generation. The main difference between these works is the source code encoding methods. Iyer et al.~\cite{DBLP:conf/acl/IyerKCZ16} directly modeled the source code as a sequence of tokens, while Hu et al.~\cite{DBLP:conf/iwpc/HuLXLJ18} used the traversal sequence of Abstract Syntax Tree (AST) tokens as the model input. LeClair et al.~\cite{DBLP:conf/icse/LeClairJM19} integrated previous work and used two different ways to represent source code. By virtue of the naturalness of the source code~\cite{DBLP:journals/cacm/HindleBGS16, DBLP:journals/csur/AllamanisBDS18}, these neural models can mine patterns for generating comments from large corpora, but they only relied on source code information, such as tokens or structures of source code, to create comments.

Note that these aforementioned comment generation methods have their own advantages. The comments generated based on the manually-crafted template methods are usually fluent and informative; the IR-based methods can take advantage of tokens in comments of similar code snippets; the neural-based methods can learn the semantic connection between natural and programming languages. Although the neural-based methods have achieved the best performance~\cite{DBLP:conf/iwpc/HuLXLJ18, DBLP:conf/acl/IyerKCZ16, DBLP:conf/icse/LeClairJM19}, it tends to generate high-frequency words in comments or "lose control" sometimes. For example, according to LeClair et al.'s study~\cite{DBLP:conf/icse/LeClairJM19}, 21\% comments in the test set contain tokens with the frequency of less than 100. Conversely, only 7\% comments predicted by their proposed approach contain tokens with the frequency of less than 100. Besides, more than two thousand generated comments even do not have a normal end-of-sequence </s> token. Specifically, the comments generated by the neural model suffer a loss in readability and informativeness. This phenomenon also appears in the use of neural models in machine translation~\cite{DBLP:conf/aclnmt/KoehnK17}. Therefore, we argue that it is not enough for the neural model to generate comments only based on the source code.

Inspired by the template-based methods and IR-based methods, we assume that comments of similar code snippets can be retrieved as templates to guide the process of neural comment generation. These templates, on the one hand, provide reference examples for generating comments, and on the other hand, may contain low-frequency words related to the source code, enhancing the neural model's ability to output low-frequency words. Considering the differences between comments of similar code snippets and manually-crafted templates, we call existing similar comments as \textbf{exemplars}. Due to the strong pattern recognition capabilities of neural networks~\cite{DBLP:journals/csur/AllamanisBDS18}, we argue that encoder-decoder neural networks can be combined with traditional template-based and IR-based methods. 

Therefore, in this paper, we propose a novel comment generation framework, namely Re$^2$Com, which consists of two modules: a \textbf{Re}trieve module and a \textbf{Re}fine module. In the Retrieve module, given an input code snippet, we exploit IR techniques to retrieve the most similar code snippet from a large parallel corpus of  code snippets and their corresponding comments, and treat the comment of the similar code snippet as an exemplar. In the Refine module, we apply a novel seq2seq neural network to learn patterns for generating comments. More specifically, the encoder takes the given code snippet, the similar code snippet, and the exemplar as input, and the decoder generates the token sequence of a comment. It is worth noting that the similar code snippet retrieved may not be semantically similar to the given code snippet. With the similar   code snippet as input, we can perform a semantic comparison through a neural network and decide whether to use the exemplar based on the degree of similarity. Furthermore, we adopt the attention mechanism~\cite{DBLP:journals/corr/BahdanauCB14} to focus on the important parts of the input. In the testing phase, given a new piece of code snippet without a comment, our approach retrieves a similar code snippet and comment pair from the corpus, utilizes the trained neural model to generate an annotation, and selects tokens with the highest probability in the vocabulary as the output. In this way, we leverage the advantages of template-based and IR-based methods, and model them into the neural network to improve the performance of comment generation.

To train and evaluate our approach, we conduct experiments on a real-world Java dataset. The dataset comes from the Sourcerer repository\footnote{https://www.ics.uci.edu/~lopes/datasets/} and has been processed by LeClair et al.~\cite{DBLP:conf/icse/LeClairJM19}, including removing duplicates, dividing into training, validation, and test sets by projects. We employ the evaluation metrics BLEU score in machine translation to evaluate the generated comments and also perform a human evaluation. Experimental results show that our method performs substantially better than the IR-based method and outperforms the state-of-the-art approaches. Besides, experimental results also show that our proposed modules are orthogonal to other techniques, i.e., applying the Retrieve and Refine modules to other neural models can improve the performance of the models.

The contributions of our work are shown as follows:
\begin{itemize}
\item We propose an exemplar-based neural comment generation method, which combines traditional template-based and IR-based methods with neural methods. We use comments of similar  code snippets as exemplars to assist in generating comments.
\item We conduct extensive experiments to evaluate our approach on a large-scale dataset of Java methods. The experimental results show that our Retrieve and Refine modules  substantially improve the performance of the neural model and achieve the state-of-the-art results.
\end{itemize}

\noindent\textbf{Paper Organization}. The rest of our paper is organized as follows. Section~\ref{sec:motivation} describes motivating examples. Section~\ref{sec:method} presents our proposed method. Section~\ref{sec:experiments} and Section~\ref{sec:results} describe the experiment setup and results. Section~\ref{sec:discussion} and Section~\ref{sec:related} discuss some results and describe the related work, respectively. Finally, Section~\ref{sec:conclusion} concludes the paper and points out future directions.

\begin{figure}[!t]
\centering
\begin{minipage}{0.9\linewidth}
    \begin{subfigure}[!t]{\linewidth}
    \centering
    \includegraphics[width=0.92\textwidth]{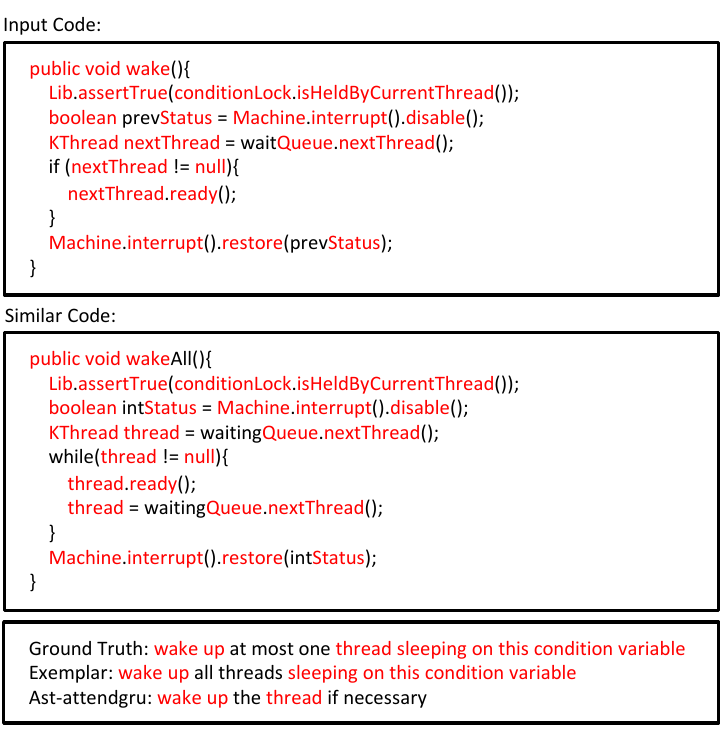}
    \caption{An example of the retrieved similar code snippet and the input code that are semantically similar.}
    \label{fig:example:a}
    \end{subfigure}
    \begin{subfigure}[!t]{\linewidth}
    \centering
    \includegraphics[width=0.92\textwidth]{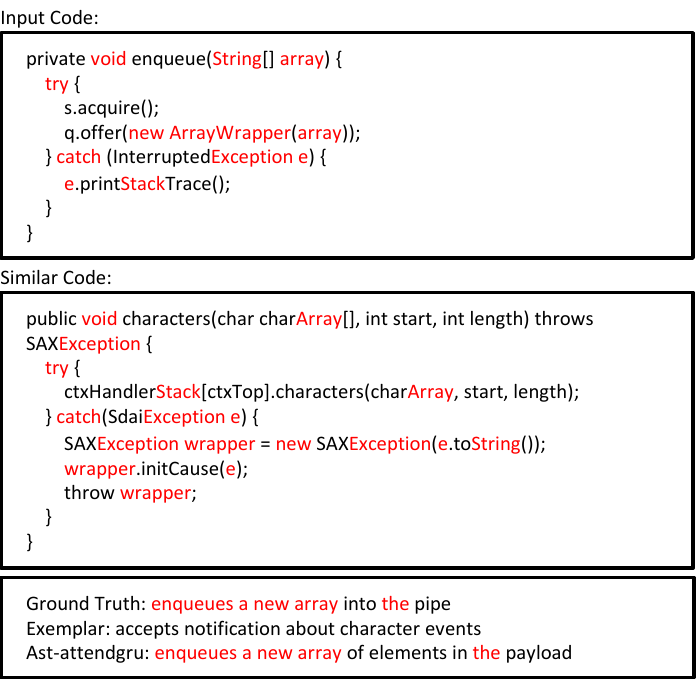}
    \caption{An example of retrieved similar code snippet and input code that are not semantically similar.}
    \label{fig:example:b}
    \end{subfigure}
\end{minipage}\vspace{-.4cm}
\caption{Examples of exemplar-based comment generation. 
Same tokens in ground truths, exemplars and predictions of ast-attendgru are marked in red. Same tokens (split on camel case) in the input code and the similar code are also marked in red.}
\label{fig:example}
\vspace{-.3cm}
\end{figure}

\section{Motivating Examples}\label{sec:motivation}
To explain why we use the comment of the retrieved similar code snippet as an exemplar to guide the neural model to generate a comment, we select two representative examples from the dataset used in the evaluation, as shown in Figure~\ref{fig:example}. The input code and the similar code are Java methods, and we also display the comments predicted by the \textbf{ast-attendgru} model~\cite{DBLP:conf/icse/LeClairJM19} for the input code. 
For the input code, we leverage the open-source search engine Lucene\footnote{https://lucene.apache.org/} to retrieve the most similar code snippet from the training corpus. The retrieval technique is based on the lexical level similarity of the source code, which will be explained in detail in Section~\ref{sec:retrieve}.

Comment generation methods based on neural networks are difficult to generate low-frequency tokens, whereas comments of similar code snippets selected based on IR-based methods may contain low-frequency tokens. For example, in Figure~\ref{fig:example:a}, we can observe that the specific phrase "sleeping on this condition variable" appears in both the ground truth and the exemplar, meaning that the input code and the similar code are semantically similar. In addition, "sleeping" is a low-frequency token in the corpus, which appears only 71 times. This is one of the reasons that the prediction of the ast-attendgru ignores the token. Although the prediction of the neural network is very close to the ground truth, it still lacks some key information in the source code. Therefore, with the exemplar as input to the neural model, the low-frequency tokens will affect the comment generation process, and the neural network will generate more informative comments.

However, it is not enough to only take an exemplar as additional input, and the similar code retrieved by search engines is not necessarily semantically similar, which is partly due to the fact that there is no real source code reuse in the corpus, and partly due to the limitations of the retrieval technique. For instance, in Figure~\ref{fig:example:b}, even though the input code and the similar code have some same tokens (Tokens in the source code are split on camel case.), they are not similar in semantics and behavior. In this case, the exemplar is unsuitable for guiding the comment generation process of neural networks. In contrast, ast-attendgru can generate a comment that is close to the ground truth without the exemplar. For this reason, we argue that it is necessary to use the similar code and the input code as the input of the neural network at the same time, and calculate the semantic similarity between the similar code and the input code through the neural model, and determine the degree of using the exemplar according to the semantic similarity. We design a novel network structure to implement this idea, the details of which will be described in Section~\ref{sec:refine}.

In view of the many discussions on the effectiveness of deep learning methods in the field of software engineering in recent years~\cite{DBLP:conf/kbse/LiuXHLXW18, Jiang2019MachineLB}, we think that our study may be a good starting point, combining traditional methods on specific tasks with deep learning methods. Previous methods applied neural networks to solve tasks in software engineering. Although specific input for specific tasks was proposed, such as AST and control flow graphs, previous researchers did not analyze the existing problems of deep learning methods, e.g., overfitting (which tends to generate high-frequency terms). Therefore, we believe that traditional methods can be modeled into neural networks to improve performance.

\begin{figure*}[t]
	\centering
	\resizebox{.9\linewidth}{!}{
	\includegraphics[width=\linewidth]{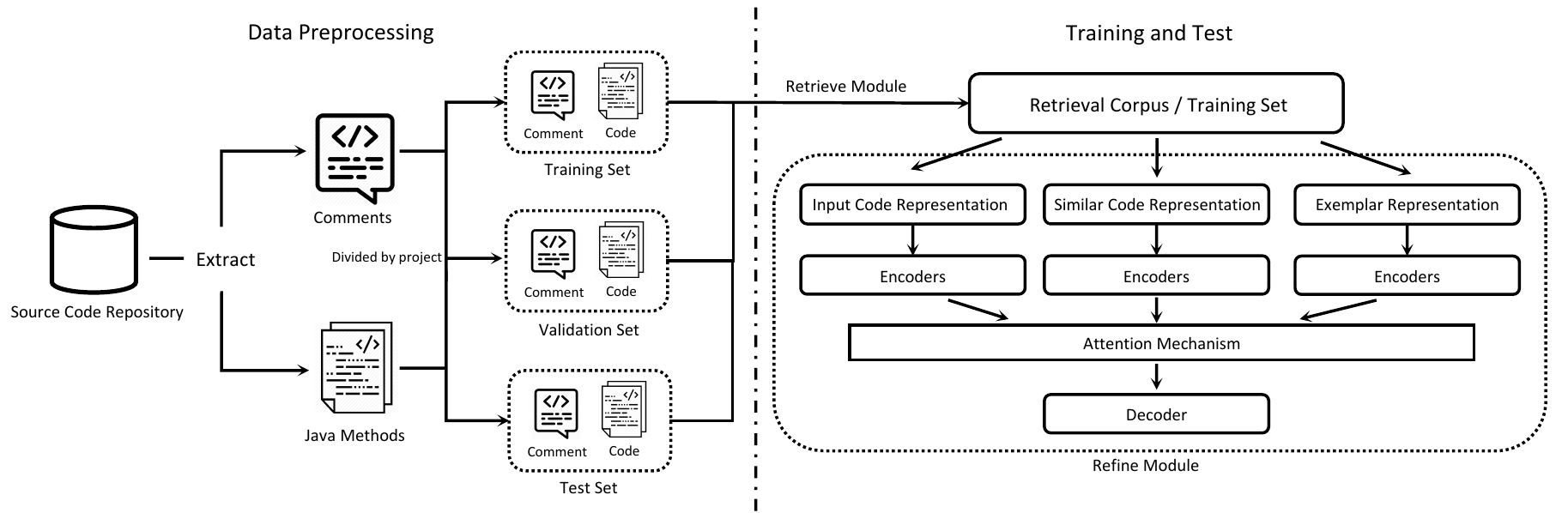}
	}\vspace{-.3cm}
	\caption{An overview of our approach for exemplar-based comment generation.}\label{fig:framework}
	\vspace{-.3cm}
\end{figure*}

\section{PROPOSED APPROACH}\label{sec:method}
In this work, we propose the exemplar-based comment generation method (Re$^2$Com), modeling traditional IR-based and template-based methods into the neural method. Different from the IR-based methods, we employ a neural network to modify the comment of the similar code to conform to the semantics of the input code. Different from the traditional template-based method, which requires manual definition of the template, we treat the comment of the retrieved code as an exemplar. Re$^2$Com consists of two parts: a Retrieve module and a Refine module. The Retrieve module uses the IR technique to explore the similar code and extract its comment from a parallel corpus, while the Refine module is a novel neural network based on a seq2seq network with an attention mechanism to generate a comment.

The overall framework is illustrated in Figure~\ref{fig:framework}. The data preprocessing step refers to the extraction, cleaning, and partition of the dataset, and the training and test step refers to the Re$^2$Com. We use a massive training set as a retrieval corpus while training the Refine module. The details of the Retrieve module are described in Section~\ref{sec:retrieve}. To take advantage of the structure information of input code, we not only use the token sequence representation of the code but also use the AST of the code as an input of the Refine module. The details of this part will be described in Section~\ref{sec:refine}.


\subsection{Retrieve Module}\label{sec:retrieve}
Considering that software reuse is widespread in software development, a similar code snippet usually has a similar comment. Furthermore, as we analyzed in Section~\ref{sec:motivation}, there are some potential problems with the previous neural-based methods. Therefore, we argue that it is beneficial for the neural network to use an exemplar as a reference when generating new comments. In practice, developers have similar experiences during software development. For example, when adding a comment to a piece of source code, they will refer to the comment of a similar code snippet. In our framework, the goal of the Retrieve module is to retrieve similar code from a retrieval corpus given the input code and treat its comment as an exemplar. 

To identify which piece of code in the retrieval corpus is most similar to the input code, we need to define and calculate the similarity between two pieces of code snippets. In this work, we chose the similarity of the lexical level of the source code to measure the code similarity, which was inspired by~\cite{DBLP:conf/kbse/LiuXHLXW18, DBLP:journals/tse/KamiyaKI02, DBLP:conf/icse/SajnaniSSRL16}. Specifically, we adopt \textit{BM25} as the similarity evaluation metric, which is a bag-of-words retrieval function to estimate the relevance of documents to a given search query in IR. Given a query and a document, based on TF-IDF, the \textit{BM25} scoring function calculates the term frequency in the document of each keyword in the query and multiplies it by the inverse document frequency of this term. The more relevant two documents are, the higher the value of \textit{BM25} score is.

We leverage the open-source search engine \textit{Lucene} to build our Retrieve module. Since the
size of the training set is quite large (over 1.9M), we use it as the retrieval corpus, i.e., given the input code snippet, we search for the most similar code from the training set. The Retrieve module contains two parts, creating the index and searching. We first tokenize the source code and comments, using \textit{WhitespaceAnalyzer} in Lucene. Then we process each code and comment pair into a document, add it to the index library, and store it on disk. In the search phase, for each query code, we get similar code sequences arranged in descending order of similarity, choose the first-ranked similar code (when training, we choose the second-ranked), and use its comment as an exemplar. We keep the default settings of \textit{BM25} in Lucene.

\subsection{Refine Module}\label{sec:refine}
Once we have an exemplar, a straightforward way is to treat it as a comment for the input code. However, due to the non-existence of software reuse or the limitation of retrieval technology, there is a certain difference between the semantic of the exemplar and the semantic of the input code. Especially, the similar code usually contains information that is inconsistent with the input code, such as different API calls and operations. Therefore, we use the exemplar as a soft-template and refine it according to the semantic difference between the source code and the similar code. Based on a widely used seq2seq neural network~\cite{DBLP:conf/nips/SutskeverVL14,DBLP:conf/emnlp/RushCW15,DBLP:journals/corr/VinyalsL15}, we design a novel network structure that can learn the semantic similarity between the input code and the similar code, refine the exemplar, and generate a comment.

\begin{figure}[t]
	\centering
	\resizebox{\linewidth}{!}{
	\includegraphics[width=0.7\linewidth]{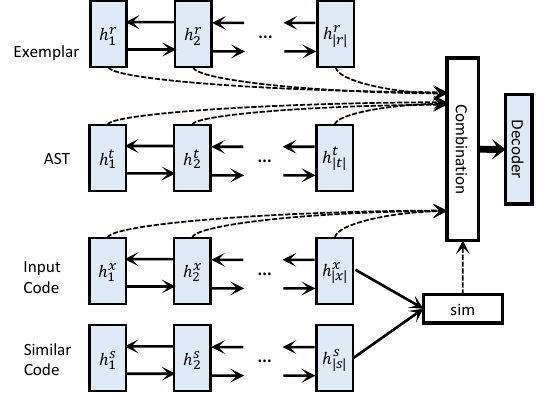}
	}
	\caption{The structure of the Refine module. The calculation details of the "sim" block and the "combination" block are Equation~\ref{eqn:sim} and Equation~\ref{eqn:combine}, respectively. The dashed lines represent information used to initialize the decoder and to calculate the context vector.}\label{fig:refine}\vspace{-.4cm}
\end{figure}

The Refine module contains three components, four encoders, a decoder, and an attention mechanism module between encoders and the decoder. Figure~\ref{fig:refine} illustrates the detailed Refine module.

\subsubsection{Encoders} 
The four encoders take a token sequence of the input code $\textbf{x}$, an AST traversal sequence of the input code $\textbf{t}$, a token sequence of the similar code $\textbf{s}$ and the exemplar $\textbf{r}$ as input, respectively. Among them, the input code $\textbf{x}$ and its AST traversal sequence $\textbf{t}$ constitute the Input Code Representation in Figure~\ref{fig:framework}. We use the Structure-based Traversal (SBT) method~\cite{DBLP:conf/iwpc/HuLXLJ18} to obtain the traversal sequence of AST to utilize the structural information of the input code. There is no difference in the structure of the four encoders. Take the input code $\textbf{x}$ as an example.

The encoder of the input code first maps the one-hot embedding of a token $w_i$ into a word embedding $x_i$:
\begin{align}
    x_i = W_e^\top w_i
\end{align}
where $W_e$ is a trainable embedding matrix. Then to leverage the contextual information, we use a bidirectional long short-term memory (LSTM)~\cite{DBLP:journals/neco/HochreiterS97} to process the sequence of the word embeddings, which is explicitly designed to avoid the long-term dependency problem. At each time step $t$, the hidden state of the forward LSTM $\overrightarrow{h}_t$ can be represented by:
\begin{align}
\overrightarrow{h}_t = \text{LSTM}(x_t, \overrightarrow{h}_{t-1})
\end{align}
The hidden states of the backward LSTM can be obtained with another LSTM.
We concatenate hidden states from two directions as the representation of the $t$-th token $h_t$ in the input code, i.e., $h_t=[\overrightarrow{h}_t;\overleftarrow{h}_{t}]$. For the traversal sequence of AST, the similar code, and the exemplar, we get their respective hidden states as $h^t$, $h^s$, and $h^r$ in the same way. We denote the hidden states of tokens of the input code as $h^x$. Note that in our experiments, we used four separate LSTMs to encode different input sequences.

Then we explore the difference between the input code and the similar code using a nonlinear $\operatorname{sigmoid}$ function to obtain a semantic similarity score $sim$:
\begin{align}
\label{eqn:sim}
    sim = \sigma(W_{sim}[h_{-1}^{x};h_{-1}^{s}])
\end{align}
where $W_{sim}$ are trainable weights, $\sigma$ stands for the $\operatorname{sigmoid}$ function, and the index "-1" stands for the last hidden state. 
According to previous work in the natural language processing community~\cite{DBLP:journals/corr/BahdanauCB14}, this structure performs well in the relevance measurement. A larger value of the score $sim$ (ranges from 0 to 1) indicates that the semantics of the input code and the retrieved code is more similar.

\subsubsection{Attention}
Attention is a component that allows the decoder to focus and place more "attention" on the relevant parts of the input
sequence as needed. It is useful to introduce this mechanism into the comment generation model. For example, when developers write comments, they use the token "get" because they notice the token "return" in the source code. Therefore, we argue that different parts of the comment are related to different parts of the source code. Similarly, after the introduction of the exemplar, the decoder can also focus on some parts of the exemplar when generating comments. The attention mechanism in the Refine module is built by the classic method of Bahdanau et al.~\cite{DBLP:journals/corr/BahdanauCB14}.

Take the attention between the target comment and the input code as an example. Specifically, for each target token $y_i$, we use the hidden state of its previous token $h_{i-1}^{'}$ to calculate the attention weights as,
\begin{align}
\widetilde{\alpha}_{ij}&=a(h_{i-1}^{'}, h_j^{x}) \\
\alpha_{ij}&=\frac{\exp\{\widetilde{\alpha}_{ij}\}}
	{\sum_k\exp\{\widetilde{\alpha}_{ik}\}}
\end{align}
where $a$ is an alignment model which scores how well the input around position $j$ and the output at position $i$ match. We use a Multi-Layer Perception (MLP)~\cite{DBLP:journals/tnn/PalM92} unit as the alignment model. Then the context vector $c_i^x$ is computed as a weighted sum of all hidden states of the input code:
\begin{align}
c_i^x = \sum_j \alpha_{ij}h_j^{x}
\end{align}
The attention weights and context vector for the exemplar $c_i^r$ and the AST traversal sequence $c_i^t$ can be computed in the same way.

\subsubsection{Decoder}
The purpose of the decoder is to generate the target comment $y$. When generating the $t$-th token in the comment, the decoder first uses an LSTM to get the $t$-th hidden state $h_t^{'}$:
\begin{align}
h_t^{'} = \text{LSTM}(h_{t-1}^{'}, y_{t-1})
\end{align}
The initial state of the decoder is a combination of the input code representation and the last hidden state of the exemplar:
\begin{align}
h_0^{'} = h^c * (1-sim) + h_{-1}^r * sim
\label{eqn:combine}
\end{align}
where $h^c$ is the feature vector of the input code, which is obtained by concatenating the last hidden state of $\textbf{x}$ and the last hidden state of $\textbf{t}$ and performing an affine transformation:
\begin{align}
h^c = W_c[h_{-1}^x;h_{-1}^t] + b_c
\end{align}
where $W_{c}$ and $b_c$ are trainable parameters. The purpose of the combination in Eqn.~\ref{eqn:combine} is that if the input code is different semantically from the similar code, that is, the value of the similarity score is low, then the decoder should pay more attention to the content of the input code. We can obtain the context vector $c_t$ in the same way:
\begin{align}
c_t^{'} = (W_c[c_{t}^x;c_{t}^t] + b_c) * (1-sim) + c_{t}^r * sim
\end{align}
Then the probability of a token $y_t$ is conditioned on the context vector $c_t$ and its previous generated tokens $y_1$,...,$y_{t-1}$, i.e.,
\begin{align}
    p(y_t|y_{t-1},...,y_1,x,t,s,r) = g(y_{t-1},h_t^{'},c_t)
\end{align}
where $g$ is a MLP layer with the \textit{softmax} activation function. 

The training objective of the Refine module is to minimize the cross-entropy:
\begin{align}
    H(y) = -\frac{1}{N}\sum_{i}^{N}\sum_{j}\text{log}p(y_j^i|y^i_{<j}, x^i, t^i, s^i, r^i)
\end{align}
where $N$ is the total number of training samples, and $y_j^i$ means the $j$-th token in the $i$-th sample. Through gradient descent optimization methods, the parameters of the Refine module can be estimated. During inference, we use a beam search~\cite{DBLP:conf/nips/SutskeverVL14} to generate comments. Specifically, the decoder generates the comment token by token from left-to-right while keeping $B$-best candidates at each time step where $B$ is the beam size.

\section{EXPERIMENT SETUP}\label{sec:experiments}
\vspace{0.1cm}\noindent{\bf Dataset. }
We evaluate our approach on the dataset provided by LeClair et al.~\cite{DBLP:conf/icse/LeClairJM19}. The original dataset comes from Lopes et al.~\cite{Lopes+Bajracharya+Ossher+Baldi:2010}, containing 5.1 million Java methods from the Sourcerer repository. Because the original dataset contains a large number of samples that are not suitable for evaluating neural models, such as repeated and auto-generated code, LeClair et al. preprocessed the data.

More specifically, they first extracted Java methods and comments from the code repository. Assuming the first sentence of the Javadoc summarizes the method's behavior~\cite{DBLP:conf/sigdoc/Kramer99}, the authors extracted the first sentence or line from the Javadoc as a comment of the method and filtered out non-English samples. Considering that the auto-generated and duplicate code (due to name changes, code cloning, etc.) will have a negative impact on neural model evaluation, the authors removed these samples using heuristic rules~\cite{Shimonaka2016IdentifyingAC} and added unique, auto-generated code to the training set to ensure that no testing was performed on these samples. After splitting camel case and underscore tokens, removing non-alpha characters, and setting to lower case, the authors divided the dataset by project into training, validation and test set, meaning that all methods in one project are grouped into one category. They argue that the preprocessing of the dataset is necessary for evaluating the performance of neural models. Without these preprocessing, the evaluation results of neural models will be inflated. For example, in the ICPC'18 paper~\cite{DBLP:conf/iwpc/HuLXLJ18}, the reported BLEU score of DeepCom is about 38, while the result on this dataset is only about 19.

\begin{figure}[!t]
\centering
\begin{minipage}{\linewidth}
    \begin{subfigure}[t]{0.5\linewidth}
    \centering
    \includegraphics[width=\textwidth]{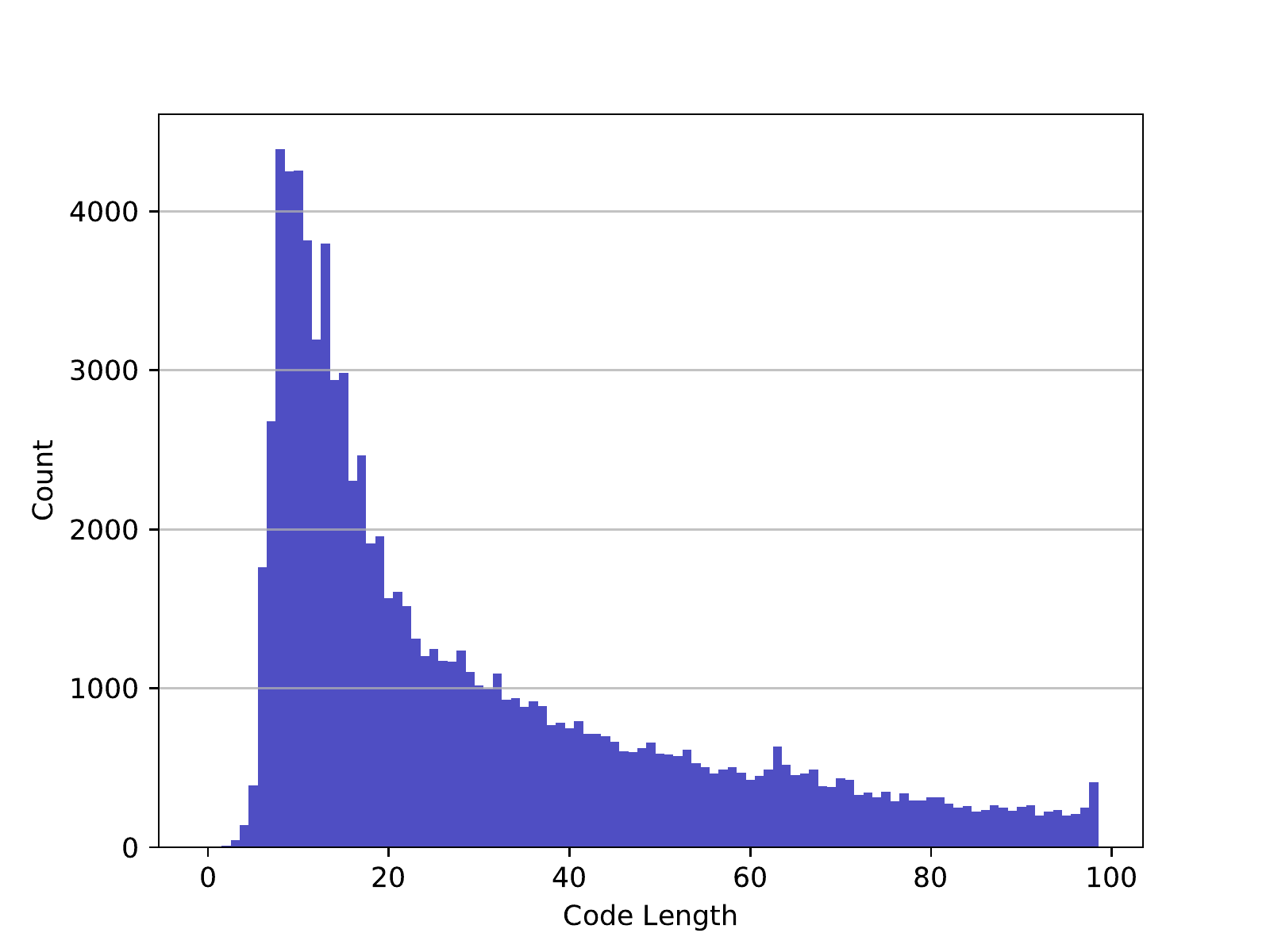}
    \caption{Code length distribution}
    \end{subfigure}%
    \begin{subfigure}[t]{0.5\linewidth}
    \centering
    \includegraphics[width=\textwidth]{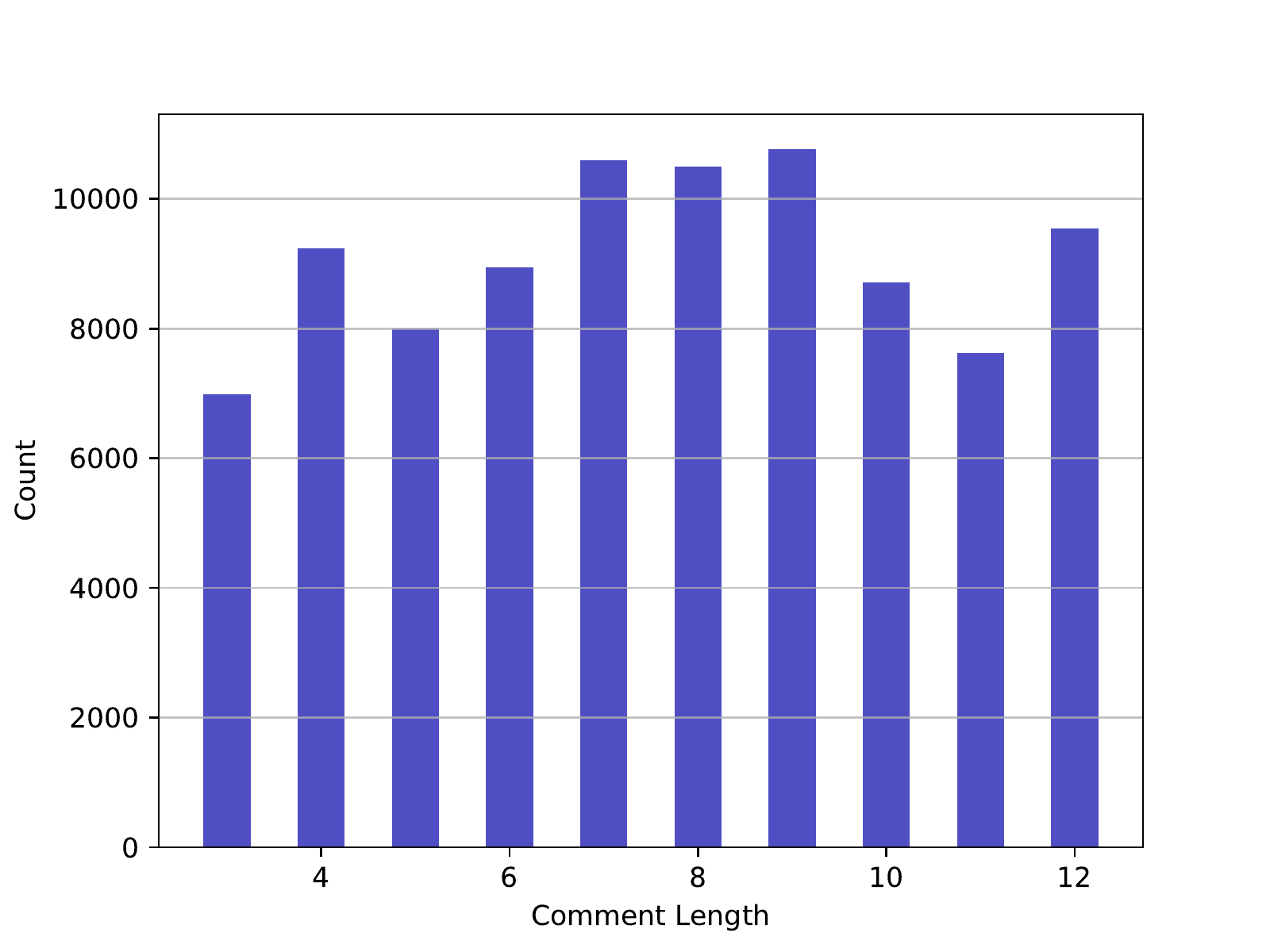}
    \caption{Comment length distribution}
\end{subfigure}
\end{minipage}
\vspace{-.2cm}\caption{Length distribution of test data}
\label{fig:distributionoflength}\vspace{-.4cm}
\end{figure}

After getting the processed dataset, the authors used the \textit{srcml}~\cite{DBLP:conf/scam/CollardDM11} tool to parse the source code into AST, and traversed the AST through the SBT~\cite{DBLP:conf/iwpc/HuLXLJ18} method to convert the AST into a token sequence. To simulate more complicated scenarios, such as missing keywords in the source code (due to poorly-written code or some scenarios with only byte code), they replaced all tokens in the source code with a <OTHER> token and got a token sequence called SBT-AO for SBT AST only. In such cases, only the structure of AST is preserved. Then the authors created two datasets to evaluate the performance of neural models.
\begin{itemize}
    \item The \textbf{standard dataset} contains three elements for each sample, a code sequence of the Java method, an SBT-AO sequence of Java method, and a comment.
    \item The \textbf{challenge dataset} contains two elements for each sample, an SBT-AO sequence of Java method, and a comment.
\end{itemize}
The challenge dataset is used to evaluate the performance of neural models when only the AST structure is available. In our experimental setup, the retrieval corpus is constructed using source code token sequence and comment pairs from the training set. For training and evaluation, we select the second-ranked (since the first-ranked similar code is itself) and top-ranked retrieved code as the similar code, respectively. The statistical results of the dataset are shown in Table~\ref{tab:dataset_stat}. Figure~\ref{fig:distributionoflength} shows the length distribution of source code and comment on the test data.

\begin{table}[t]
\small
\centering
\caption{Statistics of datasets}
\begin{tabular}{lrrr}
\toprule \
\textbf{Dataset} & \textbf{Train} & \textbf{Valid} & \textbf{Test} \\ \midrule
Count & 1,954,807 & 104,273 & 90,908 \\
Avg. tokens in comment & 7.594 & 7.710 & 7.654\\
Avg. tokens in code & 29.67 & 29.68 & 30.17\\
Avg. tokens in SBT-AO & 218.3 & 217.3 & 222.8 \\
\bottomrule 
\end{tabular}\vspace{-.4cm}
\label{tab:dataset_stat}
\end{table}

\vspace{0.1cm}\noindent{\bf Training Details. }
Our model is implemented based on the Tensorflow framework. We set token embeddings and LSTM states to 100 dimensions and 256 dimensions respectively. The out-of-vocabulary tokens are replaced by \textit{UNK}. To maximize the utilization of GPU memory, we set the batch size to 256. We choose the widely-used stochastic gradient descent to optimize all parameters with the initial learning rate of 0.2. The learning rate is decayed with a factor of 0.95 every epoch. To mitigate overfitting, we use dropout with 0.2. And to prevent exploding gradient, we clip the gradients norm by 5. According to the statistics of the dataset in Figure~\ref{fig:distributionoflength}, we limit the maximum length of the encoder LSTM to 100 and the maximum length of the decoder to 13. Training runs for about 20 epochs, and the best parameters are selected according to the performance of the validation set. During the test, the beam size $B$ is set to 5. Each experiment is run three times, and the average results are reported. We conduct our experiments on a Linux server with the NVIDIA GTX TITAN Xp GPU with 12 GB memory. 

\vspace{0.1cm}\noindent{\bf Evaluation Metrics. }
Following the previous comment generation work~\cite{DBLP:conf/acl/IyerKCZ16,DBLP:conf/iwpc/HuLXLJ18,DBLP:conf/icse/LeClairJM19}, we evaluate different approaches using the metric BLEU~\cite{DBLP:conf/acl/PapineniRWZ02}. 
BLEU measures the quality of generated comments and
can represent the human’s judgment, which calculates the similarity
between the generated comments and references. It is defined as
the geometric mean of $n$-gram matching precision scores multiplied
by a brevity penalty to prevent very short generated sentences:
\begin{align}
    BLEU = BP\cdot \exp(\sum_{n=1}^N w_n \text{log} p_n)
\end{align}
where $p_n$ is the $n$-gram matching precision scores, $N$ is set to 4 in our paper, and $BP$ is a brevity penalty to prevent very short generated sentences. BLEU score ranges from 0 to 100; the higher the score, the more the candidate correlates to the reference. This evaluation metric is also widely used in various tasks of automatic software engineering. Liu et al.~\cite{DBLP:conf/kbse/LiuXHLXW18} introduced BLEU to evaluate the quality of the generated commit message. Gu et al.~\cite{DBLP:conf/sigsoft/GuZZK16} employed it to evaluate the accuracy of the generated API sequence. Jiang et al.~\cite{DBLP:conf/kbse/JiangAM17} exploited it to evaluate the generated summaries for commit messages. Their experiments show that it is reasonable to use BLEU to evaluate the quality of comments. 
In our experiments, we report a composite BLEU score in addition to $\text{BLEU}_1$ through $\text{BLEU}_4$. 

\section{RESULTS}\label{sec:results}
To evaluate our approach, in this section, we will answer the following research questions:
\begin{itemize}
    \item RQ1: How does the Re$^2$Com perform compared to the state-of-the-art neural models?
    \item RQ2: How effective is the exemplar to all neural models?
    \item RQ3: How does the Re$^2$Com perform compared to the IR methods?
\end{itemize}
In the challenge dataset, all models have no source code tokens as input, which is an experimental scenario to evaluate the ability of the model to use limited information. In contrast, the standard dataset is close to the real-world scenario where all the information of the source code is available. Hence we evaluate our approach and all baselines on the standard dataset in this section and discuss the experimental results on the challenge dataset in the Section~\ref{sec:discussion}. 

\subsection{RQ1: Re$^2$Com vs. Neural Baselines}
\subsubsection{Baseline}
To answer this research question, we compare our approach to four state-of-the-art neural methods.
\begin{itemize}
    \item \textbf{CODE-NN}~\cite{DBLP:conf/acl/IyerKCZ16} is the first deep learning model and the first end-to-end encoder-decoder framework for comment generation task. It encodes the source code sequence into token embeddings, then uses an LSTM as a decoder to generate comments, and employs the attention mechanism to introduce information on the encoder side. Note that CODE-NN only uses token embedding as the encoder, not the LSTM.
    \item \textbf{attendgru}~\cite{DBLP:conf/icse/LeClairJM19} is a standard attentional seq2seq model, where the encoder and the decoder are both gated recurrent unit (GRU). GRU is similar to LSTM and is a variant of RNN. For a fair comparison, we replace GRU with LSTM in the model. Hence, the difference between this model and CODE-NN is whether the encoder uses an RNN.
    \item \textbf{ast-attendgru}~\cite{DBLP:conf/icse/LeClairJM19} is also an attentional seq2seq model. Different from attendgru, it introduces the structure information of the source code and uses a new encoder to process the traversal sequence of AST. It concatenates the information from the two encoders as input to the decoder and generates comments. In our experiments, we used LSTMs as encoders for a fair comparison.
    \item \textbf{DeepCom}~\cite{DBLP:conf/iwpc/HuLXLJ18} is a seq2seq model that uses LSTMs as the encoder and the decoder, and also utilizes the attention mechanism. It is the first comment generation model using AST's traversal sequence as input and proposed the traversal method SBT.
\end{itemize}
We set the embedding size and LSTM states of all baselines to 256 dimensions, which can ensure that the number of Re$^2$Com's parameters is less than the number of baselines' parameters. As in the ICSE'19 paper~\cite{DBLP:conf/icse/LeClairJM19}, we did not compare our model with other baselines in the field of natural language processing, because the tricks introduced by those models would make it difficult to compare exactly which part of the model played a key role. Compared with the above four baselines, it can not only explain the effectiveness of our model but also show that the components of our model are helpful. 

\begin{table}[t]
\small
\centering
\caption{The performance of our model compared with neural baselines.}
\vspace{-0.2cm}\begin{tabular}{l|rrrrrr}
\toprule 
\textbf{Methods} & \textbf{Params} & \textbf{B} & \textbf{B1} & \textbf{B2} & \textbf{B3} & \textbf{B4}\\ \midrule
CODE-NN & 36.3M & 12.54 & 32.23 & 14.71 & 8.558 & 6.090 \\
DeepCom & 37.9M & 14.21 & 31.88 & 16.02 & 10.10 & 7.491  \\
attendgru & 37.7M & 19.42 & 39.00 & 22.02 & 14.87 & 11.27  \\
ast-attendgru & 39.7M & 19.67 & 39.32 & 22.19 & 14.98 & 11.42 \\
Re$^2$Com & 28.4M & \textbf{24.42} & \textbf{41.69} & \textbf{25.78} & \textbf{19.70} & \textbf{16.79} \\
\bottomrule 
\end{tabular}
\label{tab:neuralbase}\vspace{-.4cm}
\end{table}

\subsubsection{Results}
We calculate the gap between the comments generated by different methods and the ground truth. The experimental results are shown in Table~\ref{tab:neuralbase}. The BLEU scores of the best baseline ast-attendgru are comparable to those reported in the study~\cite{DBLP:conf/icse/LeClairJM19}, although we made some modifications to their encoders. This result shows that the performance difference of an LSTM and a GRU on this task is very limited. Although ast-attendgru introduces structural information of the source code compared to attendgru, it does not substantially improve the results. The phenomenon also appears in the paper ~\cite{DBLP:conf/icse/LeClairJM19}, explaining that after excluding custom identifiers in the AST, the structural information of the source code has limited help in generating comments, and the information contained in the token sequence of source code is sufficient. The performance of DeepCom is much lower than the results in~\cite{DBLP:conf/iwpc/HuLXLJ18}, indicating that their data preprocessing has potential problems. The auto-generated and duplicate code has a great negative impact on the experimental results. The similar conclusion was reached in the Allamanis's paper~\cite{DBLP:conf/oopsla/Allamanis19}. The difference between DeepCom and attendgru is only in the input information, while the former is about 5 points worse than the latter. One reason is that the AST's traversal sequence processed by DeepCom is about 7 times longer than the token sequence processed by attendgru, which might contain more useless and redundant information. Koehn and Knowles~\cite{DBLP:conf/aclnmt/KoehnK17} found that encoder-decoder frameworks have low generation quality on very long sentences. From Table~\ref{tab:neuralbase}, we also notice that CODE-NN performs worst compared with other methods since it does not use an RNN to process the token sequence, which makes it unable to grasp the semantic information of the source code context. From the results, we can also observe that BLEU$_1$ to BLEU$_4$ are in descending order. BLEU$_1$ is very high compared to BLEU$_4$ on all models, revealing that the matching accuracy of the 4-grams between comments generated by neural networks and gold references is slightly lower.

From the table, we can see that Re$^2$Com substantially outperforms all neural methods on the standard dataset, and improves ast-attendgru (the best baseline) by 24.15\%. In particular, the BLEU$_4$ improvement achieves by Re$^2$Com is 47.02\%, which is not only due to the similar code retrieved and the exemplar providing abundant information for comment generation but also due to the ability of the Refine module to integrate the code and the exemplar. Since the word vectors of Re$^2$Com are 100-dimensional, the number of parameters of Re$^2$Com is the smallest among all methods. However, Re$^2$Com can still achieve the highest BLEU score. In addition, we evaluate Re$^2$Com with completely random exemplars on the test set. The model achieves 14.61 BLEU score, which shows that the improved performance of Re$^2$Com is due to the exemplar.

For a code snippet in the test set, the Retrieve module averagely takes 48.91ms to retrieve the most similar code, and the Refine module takes an average of 99.27ms to generate comments. However, the best baseline ast-attendgru takes 199.1ms to generate a comment for a given sample. Therefore, our model not only improves the results, but also improves efficiency.

\begin{table}[t]
\small
\centering
\caption{Effectiveness of exemplar on all neural methods.}
\resizebox{0.48\textwidth}{!}{
\begin{tabular}{l|rrrrrr}
\toprule 
\textbf{Methods} & \textbf{Params} & \textbf{B} & \textbf{B1} & \textbf{B2} & \textbf{B3} & \textbf{B4}\\ \midrule
CODE-NN + E. & 63.0M & 14.72 & 33.14 & 16.39 & 10.65 & 8.120 \\
DeepCom + E. & 64.4M & 18.90 & 33.51 & 19.21 & 15.50 & 13.32  \\
attendgru + E. & 64.2M & 22.60 & 39.01 & 23.48 & 18.17 & 15.68  \\
ast-attendgru + E. & 66.2M & 22.81 & 39.55 & 23.77 & 18.29 & 15.74 \\
Re$^2$Com & 28.4M & \textbf{24.42} & \textbf{41.69} & \textbf{25.78} & \textbf{19.70} & \textbf{16.79} \\
\bottomrule 
\end{tabular}}
\label{tab:validE}\vspace{-.4cm}
\end{table}

\subsection{RQ2: Effectiveness of Exemplar}
We further explore whether exemplar is effective for all neural models, i.e., when the neural model becomes simple, or when the model does not have the structure information of the source code, will the exemplar still be effective? To reach a conclusion, we first add the exemplar as input on all baselines. Then we apply the similarity score (Eqn.~\ref{eqn:sim}) and the combination block (Eqn.~\ref{eqn:combine}) in our Refine module to each baseline for calculating the initial state and context vector of the decoder. Finally, we train each baseline, and the evaluation results are shown in Table~\ref{tab:validE}.

It can be seen from the experimental results that exemplar can bring stable improvement to all neural models. For all of the baseline models, their BLEU scores are increased after adding the exemplar. Observing the improvement of BLEU$_1$ to BLEU$_4$, we can find that the exemplar has the biggest improvement on BLEU$_4$ on all models. This indicates that retrieved exemplars improve the accuracy of continuous tokens in predicted comments in the neural network, which also effectively improves the quality of generated comments. 
In addition, the number of trainable parameters of Re$^2$Com is less than all the methods. Exemplar's improvement of all the models also shows that the exemplar is orthogonal to the tricks used by other deep models, and can bring independent improvements. Take ast-attendgru and DeepCom as examples. We can see from the table that there is still a certain gap between the two models after adding exemplars, which is caused by the token sequence of the source code as an input of ast-attendgru. Overall, experimental results show that the exemplar is effective for all neural models generating comments, and still, our proposed model Re$^2$Com shows the best performance. 

\subsection{RQ3: Re$^2$Com vs. IR Baselines}
\subsubsection{Baseline}
To answer this research question, we compare our approach with four IR-based baselines. 
\begin{itemize}
    \item \textbf{Retrieve Module} is a component of Re$^2$Com, whose details are described in Section~\ref{sec:retrieve}. We use the retrieved exemplar as a comment directly.
    \item \textbf{Latent Semantic Indexing} (LSI) is an IR technique to analyze the semantic relationship between terms in documents, which is used by Haiduc et al.~\cite{DBLP:conf/wcre/HaiducAMM10} to extract important terms in source code. For a given code snippet, we use LSI to retrieve the similar code from the training set and use its comment as a target. The similarity is the cosine distance of the 500-dimensional LSI vector of the code.
    \item \textbf{Vector Space Model} (VSM) represents the source code as a feature vector and is applied to some IR-based comment generation methods~\cite{DBLP:conf/wcre/HaiducAMM10,DBLP:conf/iwpc/EddyRKC13}. We represent the source code as a vector using Term Frequency-Inverse Document Frequency (TF-IDF) and use cosine similarity to retrieve the comment of the most similar code from the training set.
    \item \textbf{NNGen}~\cite{DBLP:conf/kbse/LiuXHLXW18} is an approach for producing commit messages based on nearest neighbors. It first encodes code changes into a form of "bag of words", then uses the cosine distance to select the closest $k$ code changes, and finally chooses the message of the code change with the highest BLEU score as the final result. We replace code changes with code snippets, leverage this method to generate comments, and set $k$ as 5.
\end{itemize}

\begin{table}[t]
\small
\centering
\caption{The performance of our model compared with IR baselines.}
\vspace{-.2cm}\begin{tabular}{l|rrrrr}
\toprule \
\textbf{Methods} & \textbf{B} & \textbf{B1} & \textbf{B2} & \textbf{B3} & \textbf{B4} \\ \midrule
Retrieve Module & 18.04 & 32.06 & 17.83 & 14.39 & 12.87 \\
LSI & 17.19 & 31.38 & 17.05 & 13.48 & 12.07 \\
VSM & 17.76 & 31.91 & 17.52 & 14.02 & 12.70 \\
NNGen & 18.89 & 33.48 & 18.86 & 14.99 & 13.44 \\
Re$^2$Com & \textbf{24.42} & \textbf{41.69} & \textbf{25.78} & \textbf{19.70} & \textbf{16.79} \\
\bottomrule 
\end{tabular}
\label{tab:irbase}\vspace{-.4cm}
\end{table}

\subsubsection{Results}
We evaluate the quality of comments generated by different IR-based methods, and the results are shown in Table~\ref{tab:irbase}. Although our Retrieve module achieves high BLEU scores, it does not perform as well as Re$^2$Com, which proves the importance of the Refine module. Besides, the performance of the Retrieve module is not substantially different from that of common IR-based methods, illustrating that our Retrieve module is reasonable and effective. LSI and VSM leverage different methods (LSI vectors and TF-IDF) to represent source code as vectors, but their performance is similar. NNGen chooses the comments with the highest BLEU score and thus performs better than other IR-based methods. Note that the IR-based baselines perform very well on BLEU$_4$, even surpassing the neural network-based baselines in Table~\ref{tab:neuralbase}, i.e., the IR-based methods can achieve a high matching precision score of 4-gram, indicating that these comments are informative and have a good readability. The phenomenon also explains why Re$^2$Com can improve more on BLEU$_4$. Surprisingly, IR methods outperform some neural-based methods on BLEU scores, such as CODE-NN and DeepCom, showing that in more stringent and more realistic scenarios (no duplicate and auto-generated code), neural networks are not necessarily better than IR methods. Combining the advantages of neural networks and traditional methods, our Re$^2$Com achieves the best performance.

\subsection{Human Evaluation}
Although BLEU scores can evaluate the gap between the generated comments and references, it cannot truly reflect the semantic similarity. Therefore, we perform a human evaluation to measure the quality of comments generated by NNGen, Re$^2$Com, and ast-attendgru on the standard dataset. We follow the previous work~\cite{DBLP:conf/kbse/LiuXHLXW18,DBLP:conf/kbse/Liu0T0L19,hu2019deep,DBLP:conf/acl/IyerKCZ16} to design a human evaluation, and measure three aspects, including the \textit{similarity} of generated comments and references, \textit{naturalness} (grammaticality and fluency of the generated comments) and \textit{informativeness} (the
amount of content carried over from the input code
to the generated comments, ignoring fluency of the text). Specifically, we invite 12 volunteers with 3-5 years of Java development experience and excellent English ability for 30 minutes each to evaluate the generated comments in the form of a questionnaire. Similar to~\cite{hu2019deep}, we randomly select 300 prediction results and their references from the test set (100 from NNGen, 100 from Re$^2$Com and 100 from ast-attendgru). The 300 samples are then evenly divided into six groups, with each questionnaire containing one group. We randomly list the comment pairs and the corresponding input code on the questionnaire and remove their labels to ensure that participants cannot distinguish which comment is generated by NNGen, Re$^2$Com, or ast-attendgru. Each participant is asked to rate each sample from the above three aspects. All three scores are integers, ranging from 0 to 4. Each group is evaluated by two volunteers, and the score of a pair of comments is the average of two evaluations. Participants are allowed to search the Internet for related information and unfamiliar concepts.


\begin{table}[t]
\small
\centering
\caption{The results (standard deviation in parentheses) of human evaluation}
\vspace{-0.2cm}\begin{tabular}{l|rrr}
\toprule \
\textbf{Methods} & \textbf{Informativeness} & \textbf{Naturalness} & \textbf{Similarity}\\ \midrule
NNGen & 1.555 (1.31) & 3.560 (0.70) & 1.205 (1.37) \\
ast-attendgru & 2.575 (0.93) & 3.425 (0.86) & 2.215 (1.11)    \\
Re$^2$Com & \textbf{2.930} (1.06) & \textbf{3.820} (0.64) &  \textbf{2.640} (1.29) \\
\bottomrule 
\end{tabular}
\label{tab:human}
\end{table}

The evaluation results are shown in Table~\ref{tab:human}. Re$^2$Com surpasses NNGen and ast-attendgru in three aspects. In particular, the NNGen can generate more fluent comments than the ast-attendgru, because its comments are all retrieved from the training set. The difference in standard deviation of the three methods is very small, indicating that their scores are about the same degree of concentration. Interestingly, the scores of infomativeness of all three models are higher than those of similarity, indicating that the generated comments are more relevant to the input code than to the references. Since some references contain information about the context of Java methods, such as member variables in the class, it is not possible to generate the information for all three models with only Java methods as input.
 
\section{DISCUSSION}\label{sec:discussion}
In this section, we further compare  Re$^2$Com and the best baseline ast-attendgru. Then we discuss situations where our method performs well and threats to validity.

\begin{table}[t]
\small
\centering
\caption{The number of correctly generated low-frequency tokens}
\begin{tabular}{l|rrrr}
\toprule \
\textbf{Methods} & \textbf{$\leq$10} & \textbf{$\leq$20} & \textbf{$\leq$50} & \textbf{$\leq$100}\\ \midrule
Reference & 12,145 & 15,253 & 21,622 & 28,425 \\
ast-attendgru & 262 & 624 & 1,575 & 2,801  \\
Re$^2$Com & 422 & 1,093 & 2,808 & 4,886 \\
\bottomrule 
\end{tabular}\vspace{-.2cm}
\label{tab:low-freq}
\end{table}

\subsection{Performance on Low-frequency Tokens} 
94.8\% of tokens in the comment vocabulary of the standard dataset have a frequency of less than 100. As we described in Section~\ref{sec:intro} and~\ref{sec:motivation}, previous methods perform poorly on low-frequency tokens. To evaluate the results of the Re$^2$Com on low-frequency tokens, we collect all correctly generated tokens that appear in both prediction and reference on the test set, calculate the frequency of these tokens on the training set, and count the tokens with frequencies less than 10, 20, 50, and 100. We conduct the same analysis on ast-attendgru and count the number of low-frequency tokens in the reference on the test set. Table~\ref{tab:low-freq} shows the statistical results on low-frequency tokens. The results show that Re$^2$Com can predict more correct low-frequency tokens than ast-attendgru, which indicates that Re$^2$Com can tackle the problem of predicting low-frequency tokens. The ability to predict more tokens that appear less frequently also indicates that our Re$^2$Com has better generalization capabilities.

\subsection{Performance for Different Lengths}
\begin{figure}[!t]
\centering
\vspace{-.2cm}\begin{minipage}{\linewidth}
    \begin{subfigure}[t]{0.5\linewidth}
    \centering
    \includegraphics[width=\textwidth]{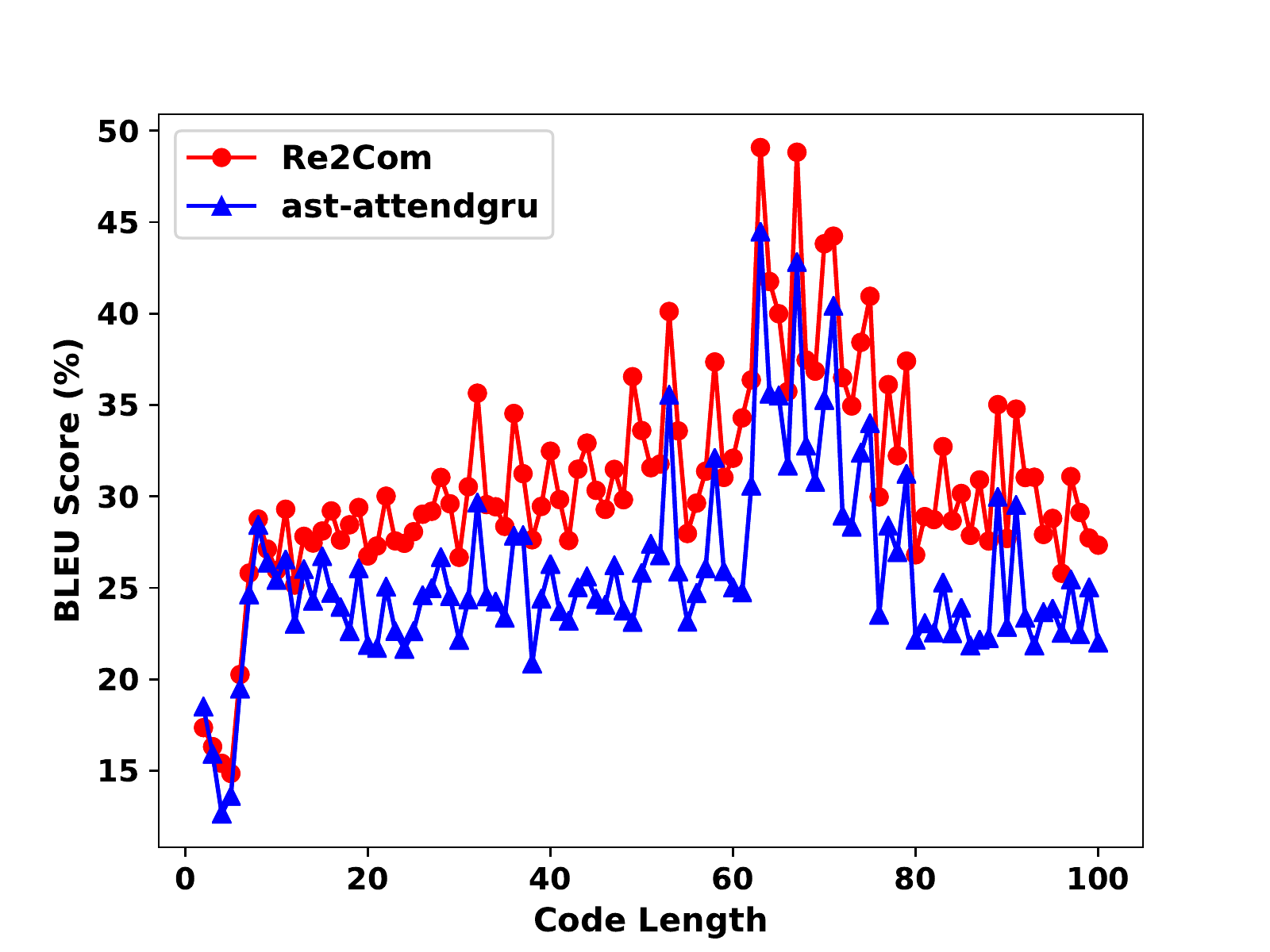}
    \end{subfigure}%
    \begin{subfigure}[t]{0.5\linewidth}
    \centering
    \includegraphics[width=\textwidth]{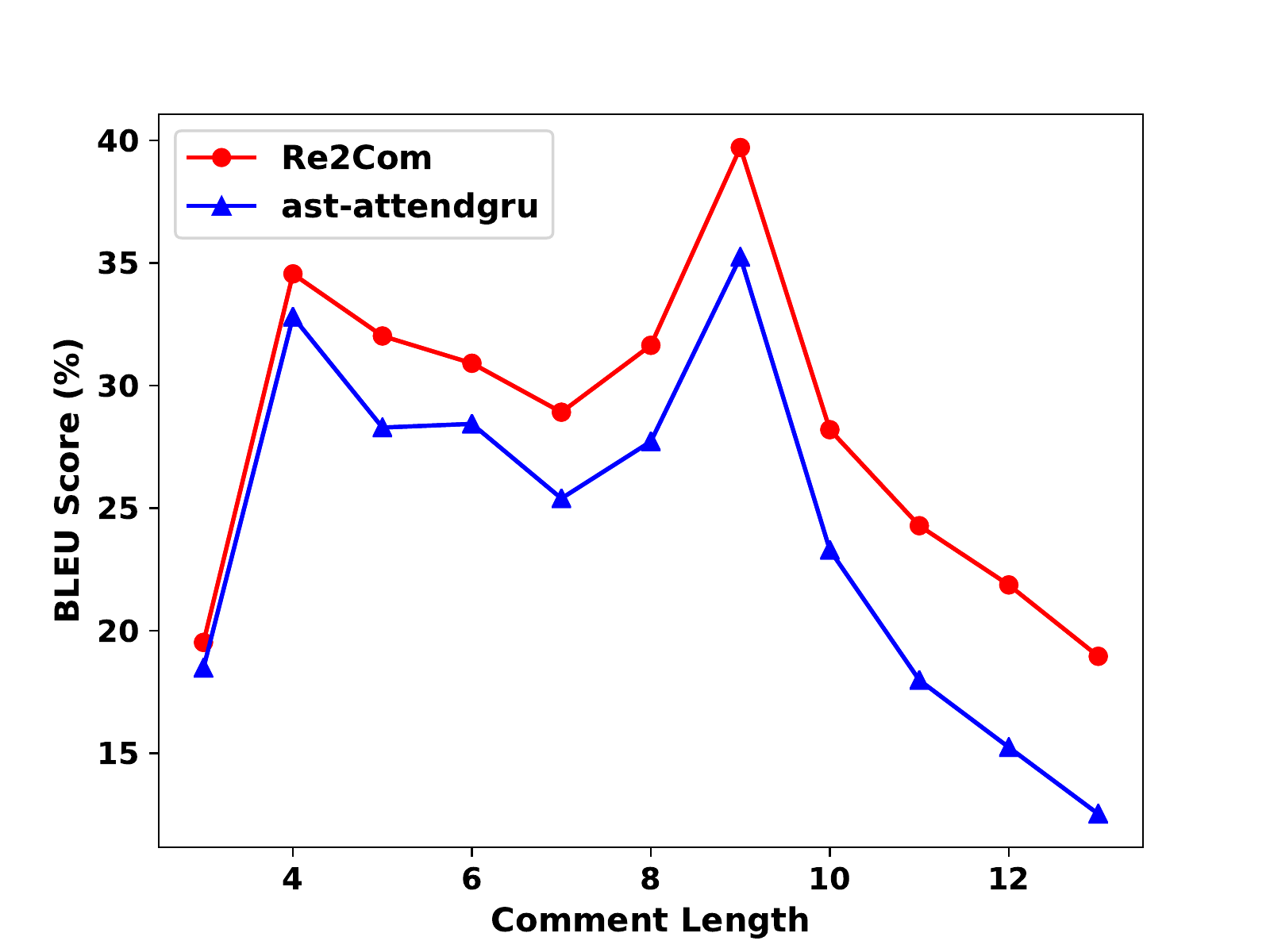}
\end{subfigure}\vspace{-.3cm}
\end{minipage}
\caption{BLEU scores for different code and comment lengths.}
\label{fig:length_res}\vspace{-.4cm}
\end{figure}
Here, we further analyze the prediction accuracy of the Re$^2$Com and the ast-attendgru on different code and comment lengths. We calculate the BLEU score for each sample on the test set and then average the scores by length. Figure~\ref{fig:length_res} shows the evaluation results. From the figures, we can observe that the Re$^2$Com outperforms the ast-attendgru with different code and comment lengths. When the code and comments are very long, the performance of both models decreases to some extent, but Re$^2$Com is still better than ast-attendgru. The improvement of Re$^2$Com is stable on code and comments of different lengths.

\begin{table}[t]
\small
\centering
\caption{The performance of our model and ast-attendgru on the challenge dataset}
\vspace{-.2cm}\begin{tabular}{l|rrrrr}
\toprule \
\textbf{Methods} & \textbf{B} & \textbf{B1} & \textbf{B2} & \textbf{B3} & \textbf{B4} \\ \midrule
ast-attendgru & 9.334 & 25.79 & 11.05 & 6.027 & 4.418  \\
Re$^2$Com & \textbf{10.50} & \textbf{27.41} & \textbf{12.26} & \textbf{7.014} & \textbf{5.182} \\
\bottomrule 
\end{tabular}\vspace{-.3cm}
\label{tab:challenge}
\end{table}

\subsection{Performance on the Challenge Dataset}
Here, we evaluate the Re$^2$Com and the ast-attendgru on the challenge dataset. Because source code token information is not available in the dataset, we use the Retrieve module to retrieve the most similar SBT-AO and treat its comment as an exemplar. The results are shown in Table~\ref{tab:challenge}. From the table, we can see that the BLEU score of our model is improved by 12.49\% compared to ast-attendgru. However, compared to the results on the standard dataset, we find that our Re$^2$Com on the challenge dataset does not perform as well as on the standard dataset, which is due to the limitation of the Retrieve module. The Retrieve module calculates the token-level similarity, and we remove all tokens from in the code to obtain the SBT-AO (details are in Section~\ref{sec:experiments}), resulting in poor retrieval results for similar SBT-AO. Therefore, Re$^2$Com does not perform well on the challenge dataset. But in the absence of code as input, it is very difficult to achieve such an improvement, and when comparing Re$^2$Com with ast-attendgru, it still proves that the Re$^2$Com is helpful for generating comments.

\subsection{Qualitative Analysis and Visualization}

\begin{table}[t]
\centering
\scriptsize
\caption{Examples of generated comments}
\label{tab:case}
\vspace{-.2cm}\begin{tabular}{c|l}
\toprule
Case ID & Example  \\
\midrule
\multirow{5}{*}{1} &
\begin{lstlisting}
public void resume() {
    Enumeration e = actuators.elements();
    while (e.hasMoreElements()) {
        ((Actuator) (e.nextElement())).resume();
    }
    e = sensors.elements();
    while (e.hasMoreElements()) {
        ((Sensor) (e.nextElement())).resume();
    }
}
\end{lstlisting}\\
&\textbf{Human-written}: resume all actuators and sensors in this mechanism \\ 
&\textbf{NNGen}: suspend all actuators and sensors on a mechanism \\
&\textbf{ast-attendgru}: resumes all actuators\\
&\textbf{Re$^2$Com}: resume all actuators and sensors in this mechanism \\ \midrule
\multirow{5}{*}{2} &
\begin{lstlisting}
public double function(double x, double y) {
    if (y >= 0) {
        return Math.pow(x, y);
    }
    else {
        return 1/Math.pow(x, -y);
    }
} 
\end{lstlisting}\\
&\textbf{Human-written}: calculates x to the power of y\\ 
&\textbf{NNGen}: get the norm of the vector squared\\
&\textbf{ast-attendgru}: returns the function value of the x y coordinate\\
&\textbf{Re$^2$Com}: method for x to the power of y\\ \midrule
\multirow{5}{*}{3} &
\begin{lstlisting}
public boolean equals(String rawSQL) {
    return TextUtil.removeLineBreaks(rawSQL).equals(
            _singleLineText);
}
\end{lstlisting}\\
&\textbf{Human-written}: check if the current element matches a given sql string\\ 
&\textbf{NNGen}: get the value of sql text\\
&\textbf{ast-attendgru}: returns true if the given sql string is equal to the given\\
&\textbf{Re$^2$Com}: check if the current element matches a given sql string\\
\bottomrule
\end{tabular}\vspace{-.4cm}
\end{table}
We perform a qualitative analysis on the generated comments. We present three Java methods with its comments from the test set and the comments generated by different methods, as shown in Table~\ref{tab:case}. We can see from the table that, thanks to the useful information provided by exemplars, the comments generated by the Re$^2$Com and the human-written comments are very close in semantics, and the Re$^2$Com performs better than other methods. 

\subsection{Threats to Validity}
One threat to validity is that we only evaluated our framework on a Java dataset. Although Java may not be representative of all
programming languages, the dataset is large and safe enough to show that our model is effective. Besides, the Re$^2$Com can be easily applied to comment generation for other programming languages.

The second threat to validity is our human evaluation. We cannot guarantee that each score assigned to every comment pair is fair. To mitigate this threat, we evaluate each comment pair by two human evaluators, and we use the average score of the two evaluators as the final score.

Another threat to validity is that the Retrieve module uses the lexical-level similarity of the source code, which may cause the code retrieved by the module to be semantically dissimilar. We recommend increasing the scale of the retrieval corpus to avoid this threat. However, in the Re$^2$Com, we introduce the Refine module to calculate the semantic similarity and decide whether to use the exemplar based on the similarity score.

\section{RELATED WORK}\label{sec:related}

\noindent{\bf Code Summarization.}
Automatic comment generation approaches vary from manually-crafted templates~\cite{DBLP:conf/kbse/SridharaHMPV10,DBLP:conf/iwpc/MorenoASMPV13,DBLP:journals/tse/McBurneyM16}, IR~\cite{DBLP:conf/wcre/HaiducAMM10, DBLP:conf/iwpc/EddyRKC13, DBLP:conf/wcre/WongLT15, DBLP:conf/kbse/WongYT13} to neural models~\cite{DBLP:conf/acl/IyerKCZ16, DBLP:conf/iwpc/HuLXLJ18, DBLP:conf/icse/LeClairJM19}. 

Comment generation based on manually-craft templates was one of the common methods for generating comments. Sridhara et al.~\cite{DBLP:conf/kbse/SridharaHMPV10} developed the Software Word Usage Model (SWUM) to capture the occurrences of terms in source code and their linguistic and structural relationships and then defined different templates for different semantic segments in source code to generate readable natural language. Moreno et al.~\cite{DBLP:conf/iwpc/MorenoASMPV13} defined heuristic rules to select relevant information in the source code, and then divided the comments into four parts, and defined different text templates for each part to generate natural language descriptions. McBurney et al.~\cite{DBLP:journals/tse/McBurneyM16} also used the SWUM model to extract the keywords in the Java method, employed the PageRank algorithm to select the important methods in the given method's context, and used a template-based text generation system to generate comments. These frameworks have achieved good results on Java classes and methods.

IR techniques have been widely used in comment generation task. 
Haiduc et al.~\cite{DBLP:conf/wcre/HaiducAMM10} used two IR techniques, Vector Space Model and Latent Semantic Indexing, to retrieve relevant terms from a software corpus, and then organized these terms into comments. Eddy et al.~\cite{DBLP:conf/iwpc/EddyRKC13} used hierarchical PAM, a probabilistic model that selected relevant terms from the corpus and included them to the comments.
Unlike the first two research works, Wong et al.~\cite{DBLP:conf/kbse/WongYT13} proposed that code snippets and their descriptions on the Q\&A sites can be used to generate comments for a piece of code. They used a token-based code clone detection tool \textit{SIM} to detect similar code snippets and used their comments as target comments. Wong et al.~\cite{DBLP:conf/wcre/WongLT15} further thought that the resources of the Q\&A sites were limited and proposed to use token-based code clone detection tools to retrieve similar code snippets from GitHub and leverage the information obtained from their comments to generate comments.

Recently many neural networks have been proposed for comment generation. With large-scale corpora for training, neural-based approaches quickly became state-of-the-art models on this task. Iyer et al.~\cite{DBLP:conf/acl/IyerKCZ16} first introduced the seq2seq model from neural machine translation into comment generation, whose encoder is the token embedding and decoder is an LSTM. Their model outperforms traditional methods on C\# and SQL summaries. Inspired by the difference between natural language and programming language, Hu et al.~\cite{DBLP:conf/iwpc/HuLXLJ18} proposed a neural model named DeepCom to capture the structural information of source code. They proposed a 
structure-based traversal method, using one LSTM to process the AST's traversal sequence, and the other LSTM to generate comments for Java methods. LeClair et al.~\cite{DBLP:conf/icse/LeClairJM19} proposed a neural method to predict the comment by combining the sequence information and structure information of the source code with two GRU encoders. In addition, they reconstructed the benchmark dataset for this task, removed duplicate and auto-generated code in the dataset, and divided the dataset into training, validation, and test by project.

Our proposed Re$^2$Com combines the advantages of the three methods, retrieves a similar code snippet from the training set, and uses its comment as the exemplar to guide the neural model for comment generation, improving performance over baselines.

\vspace{0.1cm}\noindent{\bf Code Clone Detection. }
Code clone detection that measures code similarity is a common program comprehension task in software engineering. Existing researches mainly measure the similarity between code representation varying from lexical~\cite{DBLP:journals/tse/KamiyaKI02,DBLP:conf/icse/SajnaniSSRL16} to syntactical~\cite{DBLP:conf/icse/JiangMSG07} representations. Specifically, 
CCFinder~\cite{DBLP:journals/tse/KamiyaKI02} and SourcererCC~\cite{DBLP:conf/icse/SajnaniSSRL16} are code clone detection tools based on bag of tokens, while DECKARD~\cite{DBLP:conf/icse/JiangMSG07} detects code clones based on AST. Recently, deep learning models are proposed to learn the implicit similarity between code snippets~\cite{DBLP:conf/kbse/WhiteTVP16,DBLP:conf/icsm/LiFZMR17,DBLP:conf/sigsoft/ZhaoH18,DBLP:conf/icse/ZhangWZ0WL19,DBLP:conf/ijcai/WeiL17,DBLP:conf/wcre/BuchA19}. These methods use a variety of neural networks: RtNN~\cite{DBLP:conf/kbse/WhiteTVP16}, DNN~\cite{DBLP:conf/icsm/LiFZMR17,DBLP:conf/sigsoft/ZhaoH18}, ASTNN~\cite{DBLP:conf/icse/ZhangWZ0WL19} and AST-based RNN~\cite{DBLP:conf/ijcai/WeiL17,DBLP:conf/wcre/BuchA19} to represent source code as feature vectors, and use feature vectors to calculate the similarity between source code snippets. Although we can use deep learning-based code clone detection tools to retrieve similar code snippets, these tools need to be trained on the labeled dataset. In our Retrieve module, we prefer to use a lightweight search engine and then exploit the Refine module to correct the retrieved exemplar. Therefore, we argue that the similarity at the lexical level is sufficient to find similar code snippets to assist in comment generation, and the experimental results also prove our idea.

\section{Conclusion}\label{sec:conclusion}
In this paper, we propose an exemplar-based comment generation framework named Re$^2$Com that takes advantage of three types of methods based on neural networks, templates, and IR. Our framework contains two modules, a Retrieve module for retrieving the most similar code snippet, and a Refine module that uses the comment of the similar code snippet as an exemplar to generate a target comment. In order to verify the effectiveness of our framework, we evaluated the Re$^2$Com on a large-scale Java method dataset. The experimental results show that the Re$^2$Com has a substantial improvement over the neural-based baselines and the IR-based baselines. Further analysis of the experimental results shows that the Re$^2$Com performs well not only on low-frequency tokens but also on code and comments of different lengths. In future work, we plan to explore the impact of more complex code retrieval techniques on the Re$^2$Com.

\begin{acks}
This research is supported by the National Key R\&D Program under Grant No. 2018YFB1003904, the National Natural Science Foundation of China under Grant Nos. 61832009, 61620106007 and 61751210, and the Australian Research Council's Discovery Early Career Researcher Award (DECRA) funding scheme (DE200100021).
\end{acks}

\bibliographystyle{ACM-Reference-Format}
\bibliography{sample-ref}

\end{document}